\documentclass[prd,twocolumn,floatfix,nofootinbib]{revtex4}

\usepackage{amssymb}
\usepackage[dvips]{graphicx}

\begin{document}

\title{Systematics of Staggered Fermion Spectral Properties
and Topology}

\author{Kit Yan Wong}
\affiliation{Simon Fraser University, Department of Physics, 
8888 University Drive, Burnaby, British Columbia, Canada V5A 1S6}

\author{R. M. Woloshyn}
\affiliation{TRIUMF, 4004 Wesbrook Mall, Vancouver,
British Columbia, Canada V6T 2A3}

\begin{abstract}
The spectral properties of a variety of improved staggered
operators are studied in quenched QCD. The systematic
dependence of the infrared eigenvalue spectrum on
i) improvement in the staggered operator, ii) improvement in the
gauge field action, iii) lattice spacing and iv) lattice volume,
is analyzed. It is observed that eigenmodes with small 
eigenvalues and  large chirality appear as
the level of improvement increases or as one approaches the
continuum limit. These eigenmodes can be identified as the
``zero modes'' which contribute to the chirality associated, 
via the index theorem, with the topology of the background
gauge field. This gives evidence
that staggered fermions are sensitive to gauge field topology.
After successfully identifying these would-be chiral zero modes,
the distribution of the remaining non-chiral modes is compared with
the predictions of Random Matrix Theory in different topological
sectors. Satisfactory agreement is obtained.
\end{abstract}

\maketitle

%%%%%%%%%%%%%%%%%%%%%%%%%%%%%%%%%%%%%%%%%%%%%%
\section{\label{sec_intro}Introduction}
%%%%%%%%%%%%%%%%%%%%%%%%%%%%%%%%%%%%%%%%%%%%%%

It is well known that spectral flow of the massless
Dirac operator ($D$) is related to the topology of the background
gauge fields. When massless fermions are coupled to Yang-Mills
fields with non-trivial topology, one or more eigenvalues of the
corresponding Dirac operator necessarily vanish. This is formally
given by the Atiyah-Singer index theorem \cite{Atiyah71}
\begin{equation}
Q= \mathrm{index}(D) \equiv n_{L}-n_{R},
\label{eq_indextheorem}
\end{equation}
where Q is the topological charge of the gauge field and
$n_{L}$($n_{R}$) are the numbers of left(right) zero eigenmodes
of the Dirac operator.

On the other hand, until recently it was believed that staggered
fermions do not feel gauge field topology \cite{Jansen04} because of 
the lack of zero eigenmode of the operator at finite lattice spacing.
This unpleasant feature of staggered fermions was further revealed
in comparisons of the infrared eigenvalue spectrum obtained
in simulations with the predictions of Random Matrix Theory (RMT)
\cite{berben98,gock99,Damgaard00}. The eigenvalue spectrum in all
topological charge sectors was found to be consistent with the
prediction of RMT for topological charge equal to
zero \cite{gock99,Damgaard00}.

In the past year this problem has been revisited and the situation
has changed. Recent calculations \cite{Follana04,Wenger04,wong04}
show that with staggered operator improvement
for staggered fermions the correct response
to QCD topology, governed by the index theorem, can be obtained.\footnote[1]
{Earlier Farchioni {\it et al.} \cite{Farch99}
suggested that in the Schwinger model topology may be visible 
in the spectrum of the staggered Dirac operator.}
In \cite{wong04} we presented some of our results in a preliminary
form. Here we give our final results along with the study of some
systematic effects not included in our earlier work.

The failure of unimproved staggered fermion actions to show the
proper topological features is due to lattice artifacts, in particular
flavour-changing interactions. This was already suggested in 
Ref. \cite{Kogut98}. The staggered quark action
describes four quark flavours in the continuum limit and the
eigenvalue spectrum has a 4-fold degeneracy in this limit.
At finite lattice spacing, the flavour-changing interactions
break the flavour symmetry and the degeneracy is lifted.
Consequently, staggered fermions do not have exact zero modes at
finite lattice spacing because the continuum chiral modes
(if there are any) are scattered on the lattice. One thus expects
staggered fermions to show better topological properties if
flavour-changing effects can be suppressed.

In addition to lattice artifacts, the spectral density
in the infrared limit also depends sensitively on the
volume of the lattice. This can be seen from the fact that,
because the topological susceptibility defined by
$\langle Q^2 \rangle /V$ is
independent of the volume $V$, the ensemble average $\langle Q^2 \rangle$
scales approximately with $V$. For example, previous studies
\cite{Bietenholz03,Zhang02} on the spectral properties
of the overlap operator have revealed this strong finite volume effect
(in contrast with the staggered Dirac operator, the overlap fermion operator
\cite{Neuberger98} has exact chiral symmetry \cite{Luscher98}
on the lattice and has true 
zero modes even at finite lattice spacing, see section
\ref{sec_propover}). In particular, it has been shown that
satisfactory agreement with the predictions of RMT can be
achieved only when the physical volume of the lattice exceeds
$(1.2fm)^{4}$. On smaller volumes the agreement is less appealing
and the sensitivity to topology disappears.

Both of these issues have been examined in this project by
studying the spectral flow in the infrared limit for a variety
of improved staggered operators. The improved staggered operators are
designed to suppress flavour-changing interactions by smoothing
out the quark-gluon interaction vertex \cite{Lepage99}.
Four different sources of systematic effects have been considered:
\begin{itemize}
\item improvement in the staggered operator
\item improvement in the gauge field action
\item lattice spacing
\item lattice volume.
\end{itemize}
Simulations are done both with the standard Wilson gluon action and the
$\mathcal{O}(a^{2}\alpha_{s})$ accurate tadpole improved
gluon action \cite{MILC98} for a range of lattice spacings and volumes.
It is observed that eigenmodes with small eigenvalues and
large chirality appear as the
level of improvement increases \cite{Follana04,Wenger04,wong04}. 
These small eigenmodes can be identified
as the chiral zero modes associated with the topology of the
background gauge fields. Our results also show the expected
strong dependence on volume. Only a few ``zero modes'' are
observed when the physical volume is small and the number
increases as the volume increases.

The paper is organized as follows. The improved staggered
operators and their properties will be reviewed in the next section.
A similar review on the overlap operator is given in section
\ref{sec_propover}. Details of the simulations are discussed
in section \ref{sec_simulation} and results will be presented in
section \ref{sec_results}. After successfully identifying the
``zero modes'', distribution of the remaining non-chiral modes
is compared with the predictions of RMT in Section \ref{sec_RMT}.

%%%%%%%%%%%%%%%%%%%%%%%%%%%%%%%%%%%%%%%%%%%%%%%%%%%%%%%%%%%
\section{\label{sec_propstag}Improved Staggered Operators}
%%%%%%%%%%%%%%%%%%%%%%%%%%%%%%%%%%%%%%%%%%%%%%%%%%%%%%%%%%%

The unimproved staggered Dirac operator is
\begin{equation}
D^{s}(x,y) = \frac{1}{2} \sum_{\mu} \eta_{\mu}(x)
          \left[ U_{\mu}(x)\delta_{x+\hat{\mu},y}-
                 U_{\mu}^{\dagger}(y)\delta_{x-\hat{\mu},y} \right],
\label{unimpaction}
\end{equation}
where $\eta_{\mu}(x)=(-1)^{x_{1}+\ldots+x_{\mu-1}}$ is the
standard staggered fermion phase. A variety of improvement schemes are
considered here. The basic improved version
is the $\mathcal{O}(a^{2})$ tree-level improved Asq operator
\cite{Orginos99}, which includes an additional 3-link Naik term
and replaces the gauge field in (\ref{unimpaction}) with
Fat7 effective links (sum of the original link and the
nearby paths, up to 7-link staples). 
Tadpole improvement of the $\mathrm{Asq}$ operator gives the
$\mathcal{O}(a^{2}\alpha_{s})$ accurate $\mathrm{Asqtad}$ operator.
This is the staggered fermion operator which has been used in recent 
dynamical fermion simulations \cite{Davies04}.

It has been found that there are alternate improvement schemes that
can suppress flavour-symmetry breaking lattice artifacts even more than
Asq improvement. One such scheme is based on unitarized Fat7 (UFat7) links. 
The usefulness of UFat7 smearing was first discussed by Lee \cite{lee}
and a staggered action which iterates this fattening procedure 
giving the improved Asq operators $(\mathrm{UFat7})^{n}\times\mathrm{Asq}$
was studied by Follana {\it{et al.}} \cite{Follana03}.
The HYP-improved operators $(\mathrm{HYP})^{n}$
\cite{Hasenfratz01} are constructed from unitarized fat links also
but only those links within the hypercubes containing the
original link are included in the fattening process.
Readers are referred to the respective papers for
further details.

The staggered Dirac operator is anti-Hermitian and since
$\{D^{s},\epsilon(x)\}=0$ where
$\epsilon(x)=(-1)^{x_{1}+\ldots+{x_{4}}}$, the eigenvalues
of $D^{s}$ are pure imaginary and come in pairs $\pm i\lambda$.
The commutation relation $\{D^{s},\epsilon(x)\}=0$ is a remnant
of the continuum global chiral symmetry on the lattice.
In addition, it is well known that $({D^{s}})^{2}$ connects only
even-even ($({D_{ee}^{s}})^{2}$) or odd-odd ($({D_{oo}^{s}})^{2}$) sites
on the lattice. In this project, we choose to compute
the eigenvalues of $({D_{ee}^{s}})^{2}$ in favour of $D^{s}$
since it is Hermitian and has real eigenvalues $\lambda^{2}$.
Note that the extra doubling of modes induced by squaring
is canceled by working only on even sites of the lattice.
To be precise, we compute the lowest 40 eigenvalues of
$({D_{ee}^{s}})^{2}$ which correspond to the lowest 40 positive
(imaginary) eigenvalues of $D^{s}$.

Before leaving this section it is useful to understand how the
low-lying non-chiral modes scale with the various systematics.
This explains the scales used in the spectral plots given later
in section \ref{sec_results}. In Ref. \cite{Kalkreuter95}, Kalkreuter
computed the complete spectrum of the unimproved staggered Dirac
operator on $12^{4}$ lattices at several lattice spacings.
Results show an accumulation of small eigenvalues with increasing
lattice spacing. In addition, the number of eigenvalues
increases when the volume of the lattice increases. Consequently,
the magnitude of the non-zero eigenvalues in the infrared limit
decreases with increasing lattice spacing and volume. On the other hand,
when lattice spacing decreases, i.e., one goes closer to the continuum
limit, an increase in the magnitude of the low-lying non-chiral modes
is expected. The same should happen as we come closer to the 
continuum by improving the action. These scaling properties will be
reflected in the choice of axes for the spectral graphs.

%%%%%%%%%%%%%%%%%%%%%%%%%%%%%%%%%%%%%%%%%%%%%%%%%%
\section{\label{sec_propover}The Overlap Operator}
%%%%%%%%%%%%%%%%%%%%%%%%%%%%%%%%%%%%%%%%%%%%%%%%%%

The massless overlap operator is given by \cite{Neuberger98}
\begin{equation}
D^{o} = 1 + \gamma_{5} \epsilon (H^{w}),
\label{eq_overlap}
\end{equation}
where $\epsilon(H^{w})$ is the matrix sign function
\begin{equation}
\epsilon(H^{w}) = \frac{H^{w}}{|H^{w}|},
\end{equation}
and $H^{w}=\gamma_{5}D^{w}$ is related to the standard Wilson
Dirac operator
\begin{eqnarray}
D_{a\alpha,b\beta}^{w}(x,y) & = &
\delta_{ab}\delta_{\alpha\beta}\delta_{x,y} \nonumber \\
 & - & \kappa \sum_{\mu} \left[
\left(1-\gamma_{\mu}\right)_{\alpha\beta} U_{\mu}(x)_{ab}
\delta_{x,y-\hat{\mu}} \right. \nonumber \\
 & + & \left. \left(1+\gamma_{\mu}\right)_{\alpha\beta}
U_{\mu}^{\dag}(y)_{ab} \delta_{x,y+\hat{\mu}} \right].
\end{eqnarray}
The hopping parameter $\kappa$ is related to the bare mass $m$ by
\begin{equation}
\kappa = \frac{1}{-2m+8}.
\end{equation}
A detail description of how $\kappa$ should be chosen is given
in \cite{Edwards98}. We use $\kappa=0.21$ or $m=1.62$, which
has been shown to be appropriate for our study of topology
\cite{Zhang02}.

The overlap operator (\ref{eq_overlap}) satisfies the Ginsparg-Wilson
relation \cite{Ginsparg82}
\begin{equation}
\{ \gamma_{5},D^{o} \} = D^{o} \gamma_{5} D^{o},
\end{equation}
which has exact chiral symmetry on the lattice. Consequently
in contrast with the staggered Dirac operator, the overlap operator
has true zero modes even at finite lattice spacing. The non-chiral
modes come in complex conjugate pairs.
Again, we use the operator ${D^{o}}^{\dag}D^{o}$ which is Hermitian
and positive definite and compute the lowest 5 eigenvalues in each
chiral sector. The matrix sign function $\epsilon(H^{w})$ is
approximated by a 14th order Zolotarev expansion \cite{Zolotarev32}
\begin{equation}
\epsilon(H^{w}) \simeq H^{w} \cdot \sum_{i=1}^{14} \frac{c_{i}}
{{H^{w}}^{\dag}H^{w} + b_{i}}
\end{equation}
with maximum errors $\sim \mathcal{O}(10^{-10})$ in the interval
[0.04,1.5].

%%%%%%%%%%%%%%%%%%%%%%%%%%%%%%%%%%%%%%%%%%%
\section{\label{sec_simulation}Simulations}
%%%%%%%%%%%%%%%%%%%%%%%%%%%%%%%%%%%%%%%%%%%

Simulations are done with both the Wilson gluon action and the
tadpole improved gauge field action for a range
of lattice spacings (or couplings) and volumes. The physical scale
is defined through the quenched string tension $\sqrt{\sigma}
\simeq 0.44GeV$. A new smearing method using unitarized Fat7
links \cite{Okiharu04} is used in the computation of the
Wilson loops which substantially reduces the statistical fluctuations
in the confining potential. Simulation parameters and the measured string
tensions $a^{2}\sigma$ are listed in Table \ref{table_parameters}.
The couplings were chosen carefully such that lattice spacings
agree between simulations using the Wilson gluon action and
the improved gauge field action. A total of 1000 configurations
were generated in each case.
\begin{table}
\centering
\begin{tabular}{|c|c|c|c|c|}
\hline
$\beta$ &  Action  & $a^{2}\sigma$ & $a$ ($fm$) & Volume $V$      \\ \hline
5.85    &  Wilson  & 0.0748(7)     & 0.123      & $10^{4}$        \\ \hline
6.0     &  Wilson  & 0.0478(5)     & 0.0981     &  $12^{4}$       \\ \hline
6.2     &  Wilson  & 0.0259(5)     & 0.0722     &  $16^{4}$       \\ \hline
8.26    & improved & 0.0724(5)     & 0.121      &  $8^{4}$,
$10^{4}$, $12^{4}$, $16^{4}$                                      \\ \hline
8.62    & improved & 0.0456(4)     & 0.0958     &  $12^{4}$       \\ \hline
9.18    & improved & 0.0246(5)     & 0.0704     &  $16^{4}$       \\ \hline
\end{tabular}
\caption{Simulation parameters and measured string tensions
$a^{2}\sigma$. The couplings are chosen such that lattice spacings
agree between simulations using the Wilson gluon action and the improved
gauge field action.}
\label{table_parameters}
\end{table}

\begin{figure}[!ht]
\begin{minipage}[c]{14pc}
\includegraphics[]
{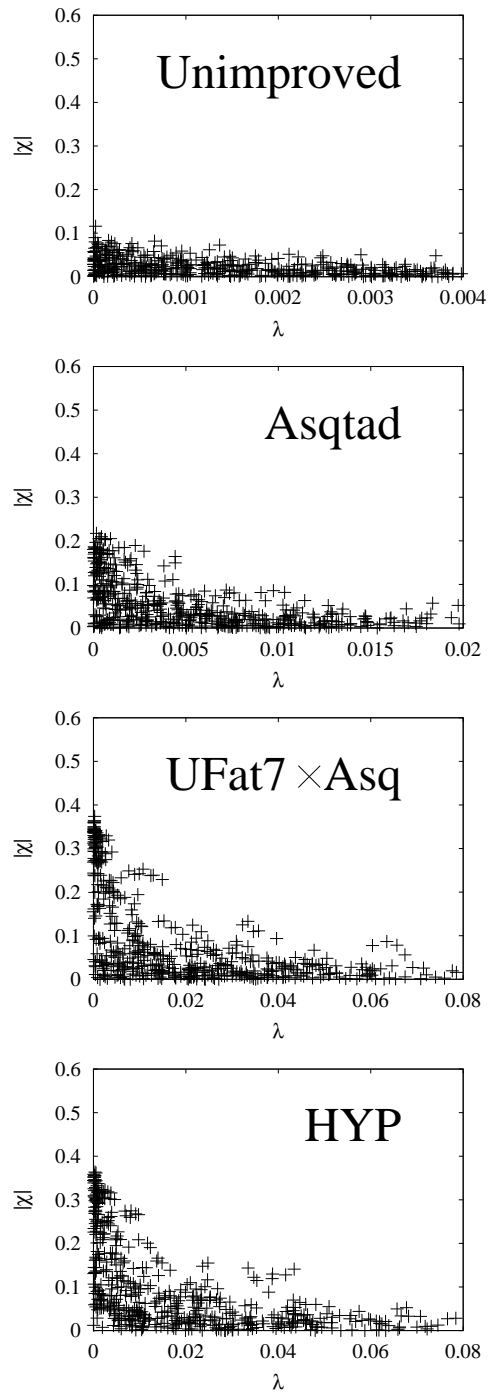}
\end{minipage}
\caption{Spectral graphs for the different operators where chirality
$|\chi|$ is plotted against the eigenvalue $\lambda$.
The ensemble is generated by the standard Wilson action at
$\beta=5.85$ ($a\simeq 0.123fm$) on $10^4$ lattice. Results are shown
for 50 configurations. Note that the scale for $\lambda$ increases
when the level of improvement increases.}
\label{fig_unimpglue_b5.85_v10_spectra}
\end{figure}

\begin{figure}[!ht]
\includegraphics[angle=-90,width=\columnwidth]
{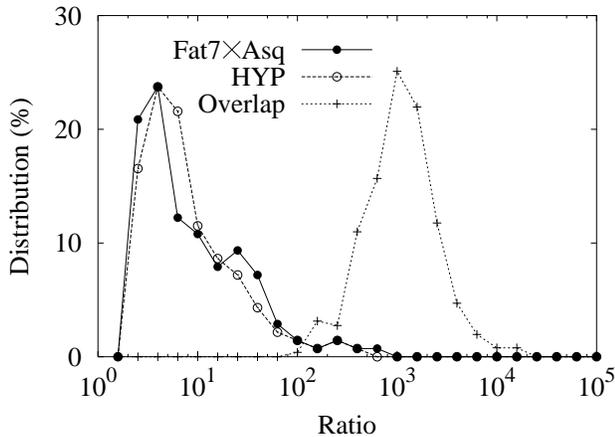}
\caption{Distribution of the ratios of eigenvalues between the smallest
non-chiral mode and the largest ``zero mode'' for configurations with
non-trivial topology. Results are shown for unimproved gauge fields
at $\beta=5.85$ ($a\simeq 0.123fm$) and lattice volume $V=10^{4}$.}
\label{fig_unimpglue_b5.85_v10_ratio}
\end{figure}

%%%%%%%%%%%%%%%%%%%%%%%%%%%%%%%%%%%%
\section{\label{sec_results}Results}
%%%%%%%%%%%%%%%%%%%%%%%%%%%%%%%%%%%%

%%%%%%%%%%%%%%%%%%%%%%%%%%%%%%%%%%%%%%%%%%%%%%%%%%%%%%%%%%%%%%%%%%
\subsection{\label{subsec_impop}Staggered Operator Improvement}
%%%%%%%%%%%%%%%%%%%%%%%%%%%%%%%%%%%%%%%%%%%%%%%%%%%%%%%%%%%%%%%%%%

We first examine the effect of staggered operator improvement
on the infrared eigenvalue spectrum. Comparisons are done
with unimproved gauge fields for $\beta=5.85$ ($a \simeq 0.123fm$)
and $V=10^{4}$. The same comparisons with improved gauge fields (same
lattice spacing and volume) will be presented in the next section.
Results are shown in Fig. \ref{fig_unimpglue_b5.85_v10_spectra}
where chirality $|\chi|$ is plotted against the eigenvalue
$\lambda$ for the different operators. Note that the chirality of 
an eigenmode is given by the expectation value of $\gamma_{5}$,
which is a 4-link operator in the staggered basis
\cite{Golterman84}. The same scale is used
for the chirality. It can be observed that the eigenvalue spectrum
depends quite sensitively on the way in which the staggered Dirac
operator is improved. Eigenmodes with small eigenvalues and relatively large
chirality appear as the level of improvement increases.
These small eigenmodes are taken to be the chiral ``zero modes''
which contribute to the index associated with the topology of the 
background gauge fields \cite{Damgaard00,Kogut98,Smit87}.
They can be identified by their separation in energy and chirality
from the rest of the spectrum (see also, Ref. \cite{Follana04}). 
Throughout this project the following criteria are used for an
eigenstate to be identified as a ``zero mode'': i) it is at least a
factor of two smaller in eigenvalue than the smallest non-chiral mode and ii)
the chirality is at least five times larger than that of the smallest
non-chiral mode.

One can see also that as the level of improvement increases,
the continuum fourfold degeneracy emerges where the scattered
eigenmodes begin to form quartets. Note that at this lattice spacing
with unimproved gauge fields the Asqtad operator is not sensitive to
the topology and lattice artifacts are still dominant.
One only starts to see a separation between the ``zero modes'' and
the non-chiral modes with further improved operators. In addition,
a large renormalization is observed for the chirality of the
would-be zero modes, which is $|\chi| \sim 0.5$
instead of unity. This large renormalization was also observed
in previous studies \cite{Kogut98,Hands90}.
Nevertheless, the ``zero modes'' can be identified
without any difficulty in most of the cases for the
$\mathrm{UFat7}\times\mathrm{Asq}$ and HYP operators.
\begin{figure}
\centering
\includegraphics[angle=-90,width=\columnwidth]
{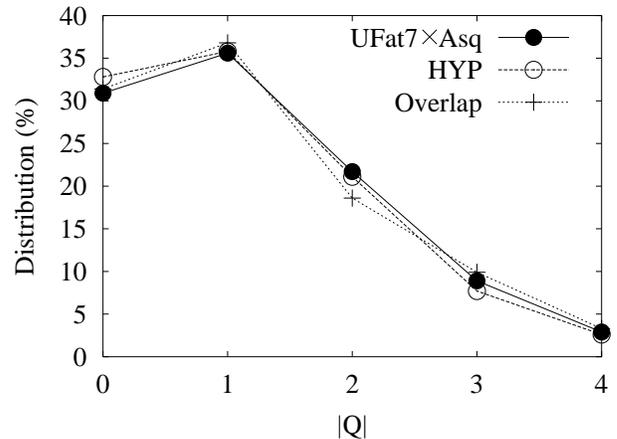}
\caption{Topological charge distributions (as a percentage of 
configurations) obtained by using
different operators (unimproved gauge fields,
$\beta=5.85$ ($a\simeq 0.123fm$) and $V=10^{4}$).}
\label{fig_unimpglue_b5.85_v10_qdist}
\end{figure}

To quantify the separation between the ``zero modes''
and the non-chiral modes, the distribution of the ratios of eigenvalues
between the smallest non-chiral mode and the largest ``zero
mode'' is plotted in Fig. \ref{fig_unimpglue_b5.85_v10_ratio}
for configurations with non-trivial topology 
(i.e., ``zero modes'' exist). We should mention that the 
configurations in Fig. \ref{fig_unimpglue_b5.85_v10_ratio} 
are different for
different operators as the topological indices obtained by using
different operators do not always agree on a configuration
by configuration basis (see next paragraph). It is possible
that for a given configuration there are no chiral modes for
one operator while they exist for the others.
For comparison results of the overlap operator
are also shown. Theoretically, the ratio is
infinite for the overlap operator because exact zero
modes exist on the lattice for overlap fermions. It is
finite here solely because of computational precision.
Results here show that the staggered operators are less
sensitive to the topology at this lattice spacing with
unimproved background gauge fields. The ratios are always
three orders of magnitude for the overlap operator but
the distribution is peaked at a ratio $\simeq 5$ for the staggered
Dirac operators. Nevertheless about 85\% of the configurations
have ratios $\ge 5$ which allows the ``zero modes'' to be identified. 
In addition our results also show that improvements using UFat7 links
or hyper-cubic blocking are equally efficient.
\begin{figure}
\begin{minipage}[c]{14pc}
\includegraphics[]
{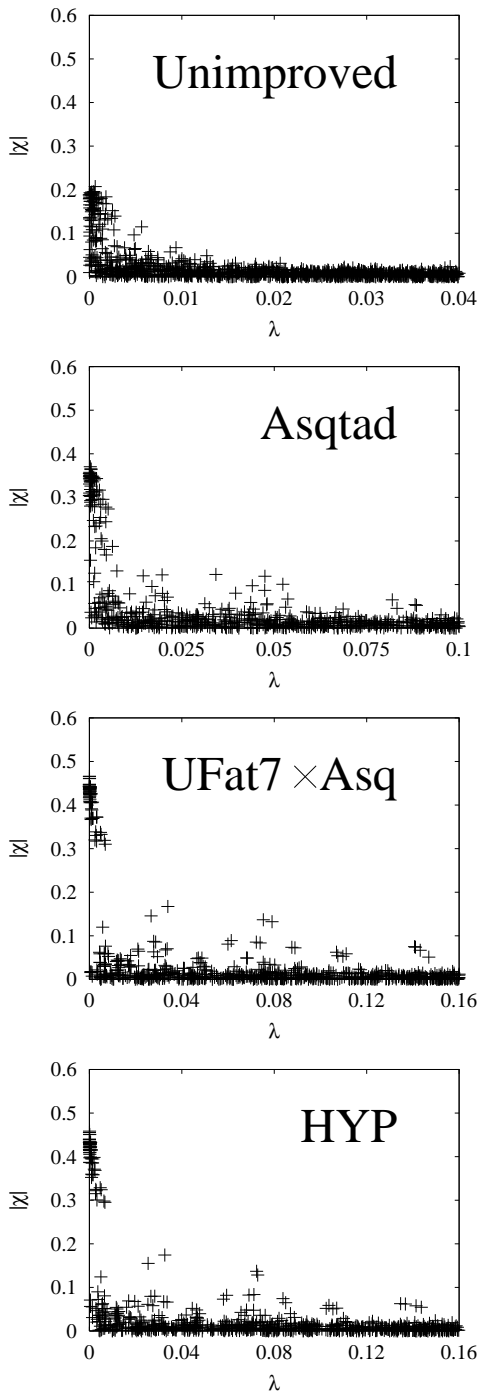}
\end{minipage}
\caption{Same as Fig. \ref{fig_unimpglue_b5.85_v10_spectra} but
with improved gauge field action ($\beta=8.26$ ($a\simeq 0.121fm$),
$V=10^{4}$). Note that the scale for $\lambda$ is larger in this
case.}
\label{fig_impglue_b8.26_v10_spectra}
\end{figure}

The identification of the topological charge index
for a given gauge filed configuration is not unique.
Because of lattice artifacts and the different ways the
staggered operator is improved, it can be expected that
the topological charge indices obtained by using different
operators do not always agree on a configuration by configuration basis.
It is only when one is close to the continuum limit that the
results agree. Even with the overlap operator previous studies
\cite{Zhang02} showed that
the topological indices do not agree on a configuration
by configuration basis when, for example, a different mass 
parameter is used in the kernel. In the present case, we
find that the topological charge indices determined by the different
operators agree about 60-70\% of the time, compared to 28\% if the values
were completely random. More explicitly, indices from different operators
are compared pairwise and the agreement is 63\% for
overlap and $\mathrm{UFat7}\times\mathrm{Asq}$,
68\% for overlap and HYP, 78\% for
$\mathrm{UFat7}\times\mathrm{Asq}$ and HYP. Note that the agreement
between $\mathrm{UFat7}\times\mathrm{Asq}$ and HYP is larger
than that obtained for 
an improved staggered operator and the overlap operator. It is then
important to check whether the charge distributions are also
different because physical observables, e.g., the topological susceptibility,
are related to the ensemble average of the topological charge.
The distributions obtained by these operators are
given in Fig. \ref{fig_unimpglue_b5.85_v10_qdist}. It can be
observed that there is no significant difference among the
results. This is important because it indicates that physics
is independent of the discretization scheme and one would expect
topological quantities obtained by these operators to agree.

%%%%%%%%%%%%%%%%%%%%%%%%%%%%%%%%%%%%%%%%%%%%%%%%%%%%%%%%%%%%%%%%%%%%%%%%
\subsection{\label{subsec_imgauge}Improvement in the Gauge Field Action}
%%%%%%%%%%%%%%%%%%%%%%%%%%%%%%%%%%%%%%%%%%%%%%%%%%%%%%%%%%%%%%%%%%%%%%%%

Lattice artifacts can be further suppressed if improvement
is also applied to the gauge field action. This can be seen
in Fig. \ref{fig_impglue_b8.26_v10_spectra} where the infrared
eigenvalue spectra of the different staggered operators are
shown for configurations generated using the tadpole
improved gauge field action. Here $\beta=8.26$
($a \sim 0.121fm$) and lattice volume is $V=10^{4}$ so that both
the lattice spacing and physical volume are very similar to those
used in the unimproved case. Results here show that
better topological properties are realized when the gauge field
action is also improved. In particular even the Asqtad operator
is sensitive to the topology at this coarse lattice spacing and
``zero modes'' can be identified unambiguously for the
$\mathrm{UFat7}\times\mathrm{Asq}$ and HYP operators.

\begin{figure}
\includegraphics[angle=-90,width=\columnwidth]
{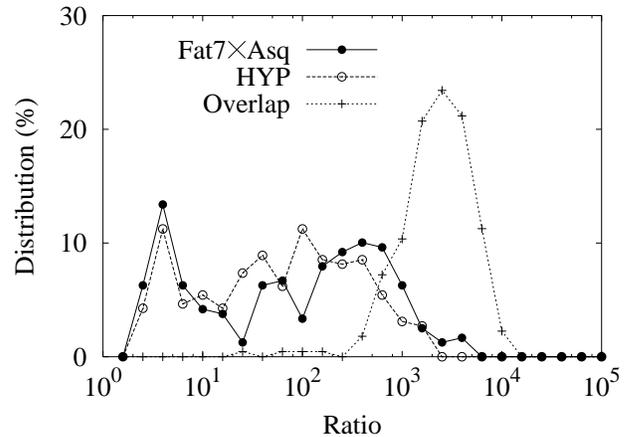}
\caption{Same as Fig. \ref{fig_unimpglue_b5.85_v10_ratio} but
with improved gauge field action ($\beta=8.26$
($a\simeq 0.121fm$), $V=10^{4}$).}
\label{fig_impglue_b8.26_v10_ratio}
\end{figure}

\begin{figure}
\centering
\includegraphics[angle=-90,width=\columnwidth]
{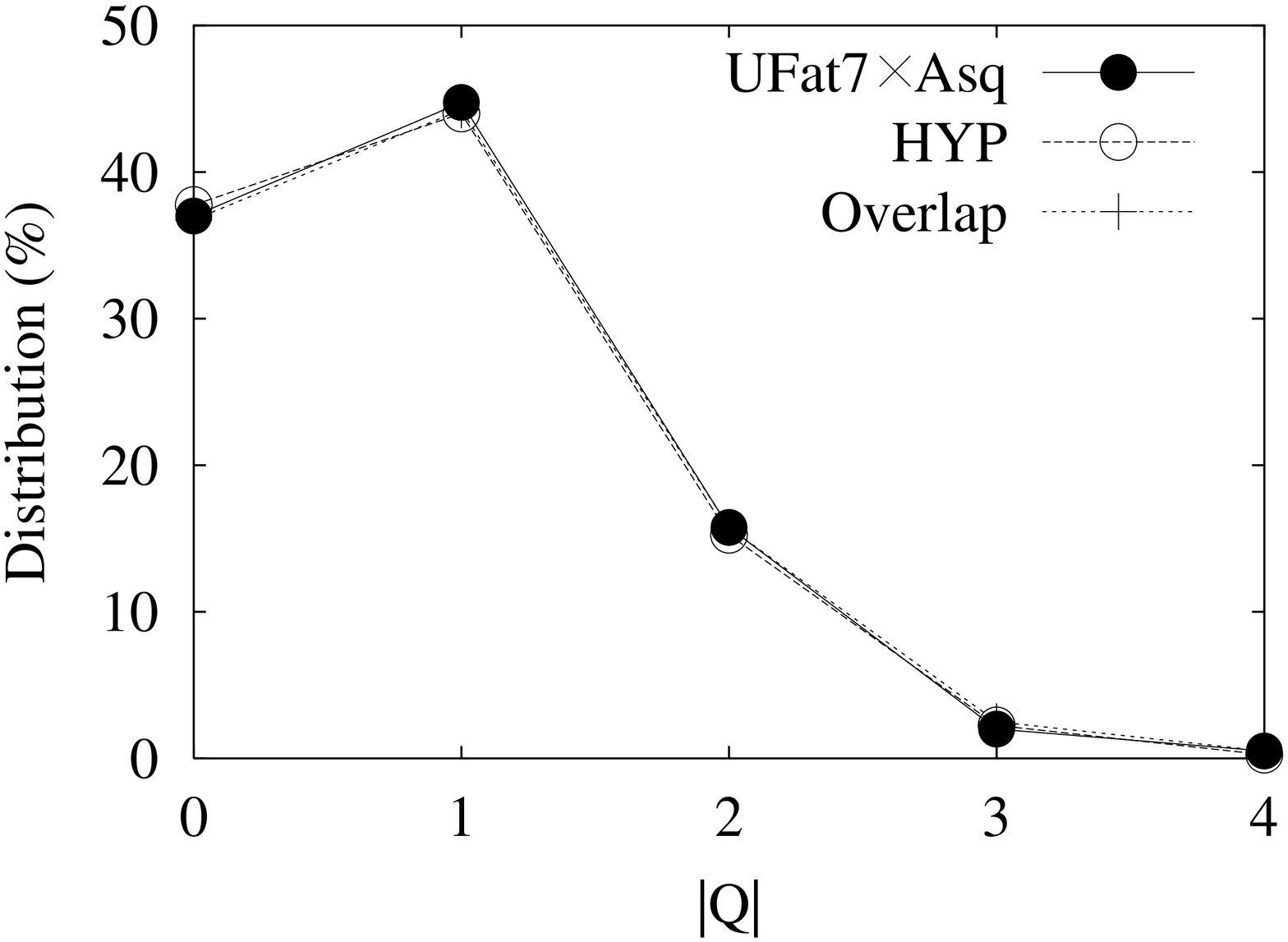}
\caption{Same as Fig. \ref{fig_unimpglue_b5.85_v10_qdist} but
with improved gauge field action ($\beta=8.26$
($a\simeq 0.121fm$), $V=10^{4}$).}
\label{fig_impglue_b8.26_v10_qdist}
\end{figure}

\begin{figure*}
\includegraphics[]
{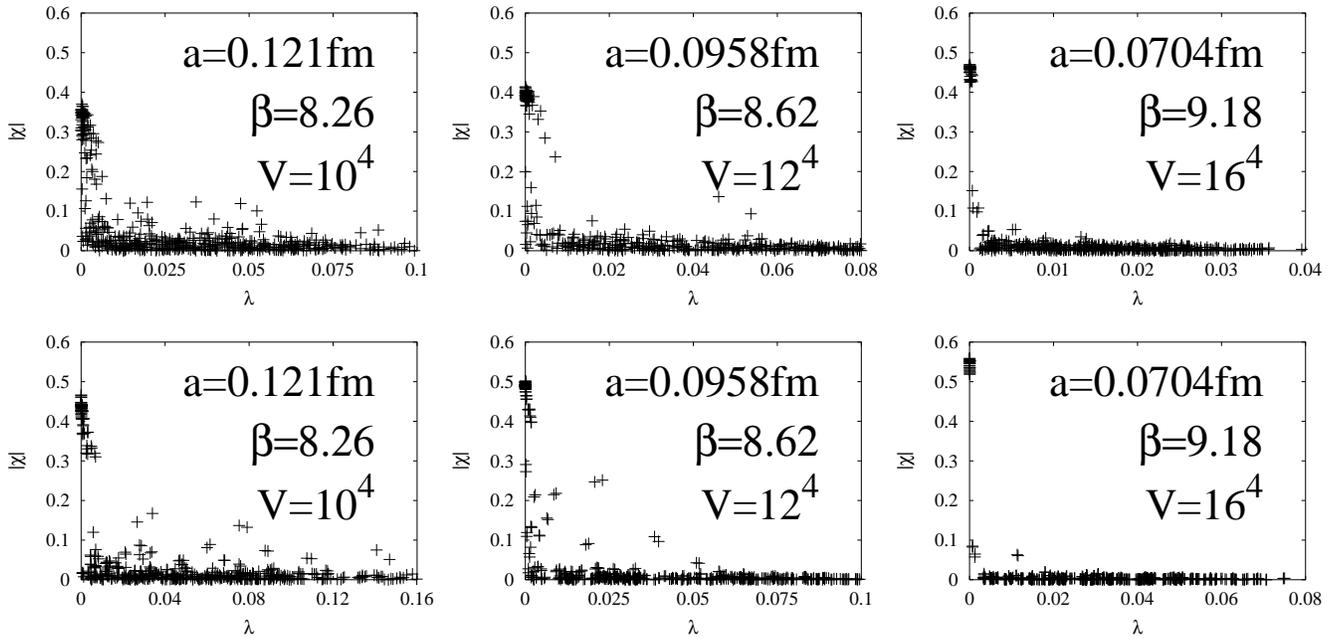}
\caption{Dependence on lattice spacing. Results are shown for the
Asqtad (top) and the $\mathrm{UFat7}\times\mathrm{Asq}$
(bottom) operators with improved gauge fields. Physical volumes
are $\sim (1.2fm)^{4}$ in all cases.}
\label{fig_beta}
\end{figure*}

To have a quantitative picture of how improvement in the gauge
field action affects the spectral flow, we again plot the
distribution of the ratios of eigenvalues between the smallest
non-chiral mode and the largest ``zero mode'' in Fig.
\ref{fig_impglue_b8.26_v10_ratio} for configurations
with non-trivial topology.
The same scale is used in Fig. \ref{fig_unimpglue_b5.85_v10_ratio}
and Fig. \ref{fig_impglue_b8.26_v10_ratio}. 
In comparison to Fig. \ref{fig_unimpglue_b5.85_v10_ratio},
it can be observed
that, for $\mathrm{UFat7}\times\mathrm{Asq}$ and HYP, 
the distribution has shifted significantly toward ratios of two or three orders
of magnitude. Overall, more than 95\% of the
configurations have ratios $\ge 5$, compared to only 85\% in the
case with unimproved gauge fields. Hence gauge field improvement
clearly increases the separation between the chiral ``zero modes'' and
the non-chiral modes. We should again emphasize here that the ratios
should be infinite theoretically for the overlap operator since exact
zero modes exist on the lattice.

We have also compared the topological indices obtained by the
different operators in this case. It is found that the agreement
increases significantly
when the gauge field action is also improved: 91\% between overlap and
$\mathrm{UFat7}\times\mathrm{Asq}$, 90\% between overlap and HYP and
96\% between $\mathrm{UFat7}\times\mathrm{Asq}$ and HYP which is again
the highest. The charge distributions are given in Fig.
\ref{fig_impglue_b8.26_v10_qdist}. It can be observed upon
comparing with Fig. \ref{fig_unimpglue_b5.85_v10_qdist} that better
agreement is obtained with improved gauge fields. Results here are
significant because they indicate that different operators do
respond the same way to the topology of the background
gauge fields when discretization errors and lattice artifacts are
reduced. In particular, results here show that the charge indices
obtained by using the staggered operators and the overlap operator,
two completely different representations of the Dirac operator on
the lattice, agree even on a configuration by configuration basis at a high
percentage as the level of improvement increases.

%%%%%%%%%%%%%%%%%%%%%%%%%%%%%%%%%%%%%%%%%%%%%%%%%%%%%%%%%%%%%
\subsection{\label{subsec_beta}Dependence on Lattice Spacing}
%%%%%%%%%%%%%%%%%%%%%%%%%%%%%%%%%%%%%%%%%%%%%%%%%%%%%%%%%%%%%

Here we examine the dependence of the infrared eigenvalue
spectrum on lattice spacing and study the spectral flow as one
approaches the continuum limit. Calculations
are done at three lattice spacings with fixed physical
volume $V \sim (1.2fm)^{4}$ (so lattice volume increases as lattice
spacing decreases, see Table \ref{table_parameters}).
Results are shown for the Asqtad and $\mathrm{UFat7}\times\mathrm{Asq}$
operators in Fig. 7. One sees that
separation between the ``zero modes'' and non-chiral modes becomes more
clear. In addition the continuum 4-fold degeneracy is better realized as one
approaches the continuum limit. Note that the chirality of the
``zero modes'' increases as the lattice spacing decreases and
it is larger for the $\mathrm{UFat7}\times\mathrm{Asq}$ operator.
This gives evidence to the fact that discretization errors and
lattice artifacts are indeed responsible for the failure of
staggered fermions to be sensitive to gauge field topology on coarse lattices.
The chiral ``zero modes'' associated with the topology of the background
gauge fields emerge as one approaches the continuum limit, a conclusion
also obtained in \cite{Wenger04}.

\begin{figure*}
\includegraphics[]
{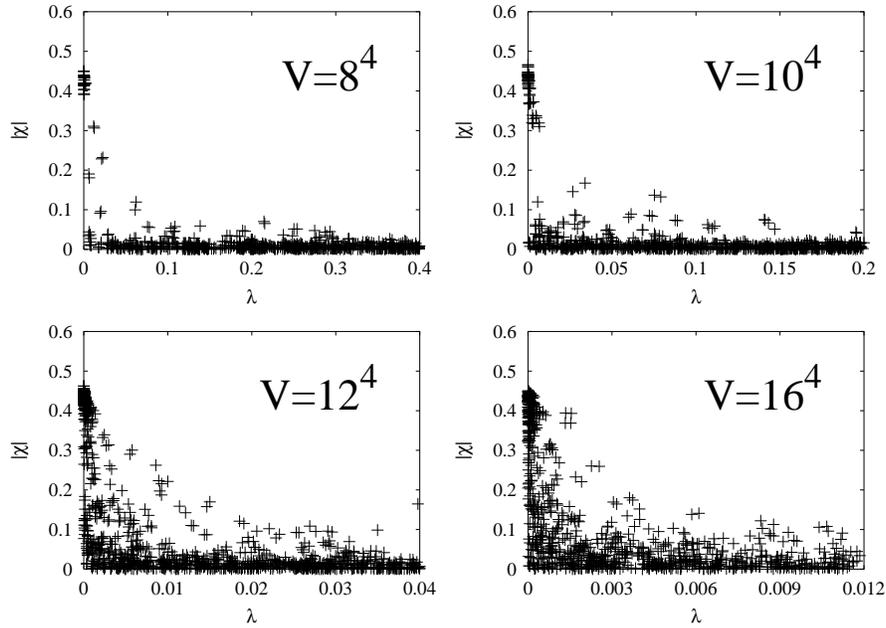}
%\begin{minipage}[t]{14pc}
\caption{Spectral graphs for the $\mathrm{UFat7}\times\mathrm{Asq}$
operator at a fixed coupling $\beta=8.26$ ($a\simeq 0.121fm$) for volumes
$V=8^{4}$, $10^{4}$, $12^{4}$ and $16^{4}$ with improved gauge fields.
The spectrum becomes noisy again for large volumes. Note that the scale
for $\lambda$ decreases with increasing volume.}
%\end{minipage}
\label{fig_vol}
\end{figure*}

Our results also show that it is necessary to use a
lattice spacing $a \lesssim 0.1fm$ for the Asqtad
operator to be sensitive to gauge field topology. This is similar to
the lattice spacings used nowadays in dynamical
simulations of QCD using the Asqtad fermion action and the tadpole
improved gauge field action \cite{Davies04}. On the other hand, further
improvement of the staggered quark action, e.g, actions with
unitarized fat-links $\mathrm{UFat7}\times\mathrm{Asq}$, may be
necessary when working on coarse lattices to ensure that continuum
physics is reproduced correctly.

%%%%%%%%%%%%%%%%%%%%%%%%%%%%%%%%%%%%%%%%%%%%%%%%%%%%%%%%%%%
\subsection{\label{subsec_vol}Dependence on Lattice Volume}
%%%%%%%%%%%%%%%%%%%%%%%%%%%%%%%%%%%%%%%%%%%%%%%%%%%%%%%%%%%

As discussed in the introduction, we expect a strong dependence
of the spectral density on the volume $V$ of the lattice
and the topological charge average $\langle Q^2 \rangle$ should scale with $V$.
In Fig. 8 the infrared eigenvalue spectrum of the
$\mathrm{UFat7}\times\mathrm{Asq}$ operator is given for
lattice volumes $V=8^{4}$, $10^{4}$, $12^{4}$ and $16^{4}$
at a fixed lattice spacing $a=0.121fm$ ($\beta=8.26$) with
improved gauge fields. Results from previous sections showed
that ``zero modes'' should be visibly separated from the
non-chiral modes at this spacing for configurations generated
by the improved gauge action. This is the case when the volume
is $\sim (1.2fm)^{4}$. A similar volume effect has been seen with 
the overlap fermion operator also \cite{Bietenholz03,Zhang02}.
A reasonable assumption is that this minimum volume effect is
related to a property of the gauge field, perhaps some minimum
volume is necessary before the topological structure is fully
formed.

As the lattice volume increases the number of eigenvalues
increases and one may expect the gap between
the zero modes and the non-chiral modes to scale as $V^{-1}$
\cite{Hands90}. The net result is that there are more low-lying
non-chiral modes as volume gets bigger. As Fig. 8 shows, a
certain volume is necessary before the would-be ``zero modes''
show up and sensitivity to topology is established. 
\begin{figure}
\centering
\includegraphics[angle=-90,width=\columnwidth]
{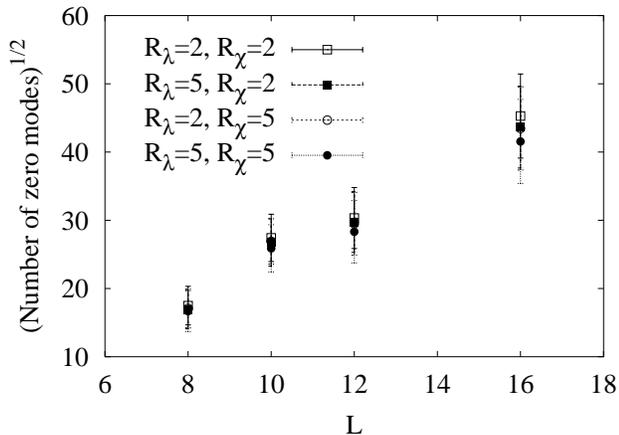}
\caption{The square-root of the total number of zero modes
of the UFat7$\times$Asq operator as a function of lattice size $L$
(lattice volume $V=L^4$). Results of using different criteria for 
identifying the zero modes are shown.}
\label{fig_num_zero}
\end{figure}
As the volume is increased even further the number of chiral ``zero modes''
increases roughly as $\sqrt{V}$ as can be inferred from 
Fig. \ref{fig_num_zero}. In the configuration average, the tail
of the chiral-mode distribution begins to merge with the low eigenvalue 
tail of the non-chiral mode distribution. However, the indentification of 
the would-be ``zero modes'' is not quite as difficult as it may appear
from Fig. 8. As seen from Fig. \ref{fig_num_zero} 
the criteria which we adopt for a  ``zero mode'', 
namely, at least a factor of 2 smaller in eigenvalue than the
smallest non-chiral mode ($R_{\lambda}=2$)
and a factor of 5 larger in chirality ($R_{\chi}=5$), are 
very robust even for the $16^4$ lattice. Imposing other values 
for these factors gives very similar results.
\begin{figure}
\centering
\includegraphics[angle=-90,width=\columnwidth]
{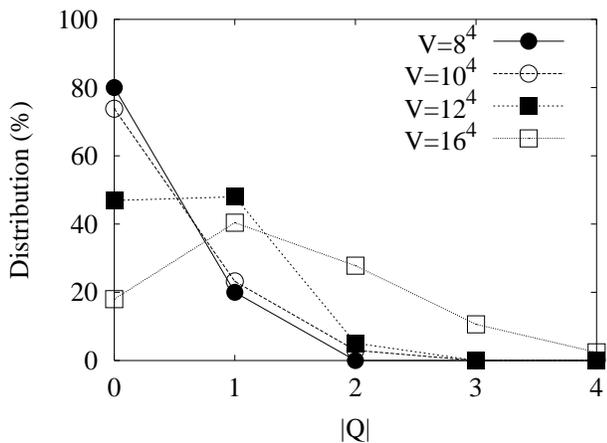}
\caption{Topological charge distribution (as a percentage of configurations)
for different lattice volumes.
Results are shown for the $\mathrm{UFat7}\times\mathrm{Asq}$
operator at $\beta=8.26$ ($a\simeq 0.121fm$) with improved gauge fields.}
\label{fig_vol_qdist}
\end{figure}

The expected increase of $\langle |Q| \rangle$ with the volume of the lattice
can be seen in Fig. \ref{fig_vol_qdist} where
the topological charge distribution is shown for different
lattice volumes. Most configurations have trivial
topological structure when the volume is small and the charge
average $\langle |Q| \rangle$ increases gradually with the volume of the lattice.

\begin{figure*}
\includegraphics[]
{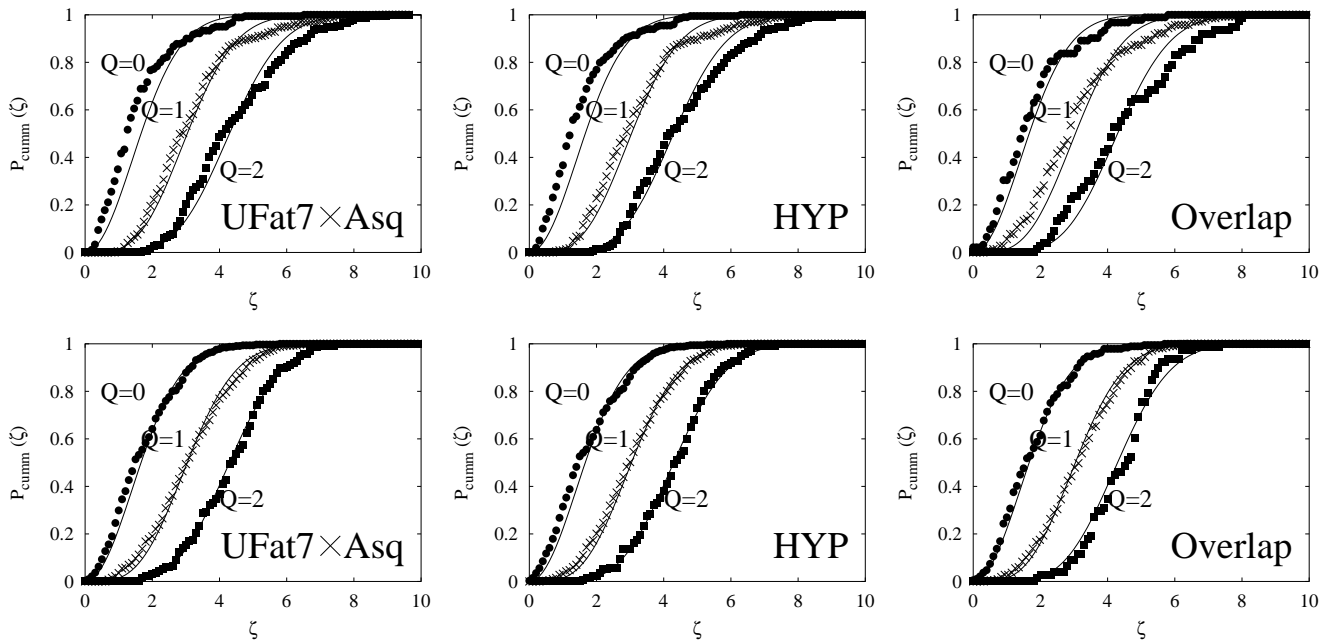}
\caption{Cumulative distribution of the smallest non-chiral
modes in different topological sectors. Results are shown
for unimproved (top) and improved (bottom) gauge fields at
$a \sim 0.12fm$, $V=10^{4}$. Solid curves are predictions
of RMT.}
\label{fig_rmt}
\end{figure*}

%%%%%%%%%%%%%%%%%%%%%%%%%%%%%%%%%%%%%%%%%%%%%
\section{\label{sec_RMT}Comparison with RMT}
%%%%%%%%%%%%%%%%%%%%%%%%%%%%%%%%%%%%%%%%%%%%%

Results from our systematic analysis give strong
indications that staggered fermions are sensitive to
gauge field topology when lattice artifacts are sufficiently suppressed.
It is then important to compare the
distribution of the non-chiral modes, after successfully
identifying the would-be zero modes, against the predictions
of RMT. In particular, we study the cumulative distribution of
the smallest non-chiral modes in different topological
sectors $P_{cumm}^{(Q)}(\zeta)$ where RMT gives the
predictions \cite{Damgaard00,Verb93}
\begin{eqnarray}
P_{cumm}^{(Q)}(\zeta) & = & \frac{\int_{0}^{\zeta} P^{(Q)}(x) dx}
{\int_{0}^{\infty} P^{(Q)}(x) dx}, \nonumber \\
P_{cumm}^{(0)}(\zeta) &
= & \frac{1}{2} \int_{0}^{\zeta} x e^{-x^{2}/4} dx, \nonumber \\
P_{cumm}^{(1)}(\zeta) &
= & \frac{1}{2} \int_{0}^{\zeta} I_{2}(x) e^{-x^{2}/4} dx, \nonumber \\
P_{cumm}^{(2)}(\zeta) &
= & \frac{1}{2} \int_{0}^{\zeta} 
\left[ I_{2}(x)^{2}-I_{1}(x)I_{3}(x) \right] e^{-x^{2}/4} dx.
\end{eqnarray}
Here, $P^{(Q)}(x)$ denotes the distribution of the smallest
non-chiral modes in charge sector $Q$ and $I_{n}(x)$ is the
modified Bessel function of the first kind (order $n$).
The variable $\zeta\equiv\lambda \Sigma V$ is related to the
infinite-volume chiral condensate $\Sigma$ which is a fitting parameter.

Calculations are done for three different operators
($(\mathrm{UFat7})\times\mathrm{Asq}$,
HYP and overlap) on two ensembles
of configurations (unimproved and improved gauge fields with
$a \sim 0.12fm$, $V=10^{4}$). These ensembles are specifically
chosen because the size of the lattice is limited by the cost in
computing the eigenvalue spectra of the overlap operator while a coarse
grid is used so that the physical volume is larger than the
critical value $(1.2fm)^{4}$. This criterion is important as
it ensures that we are in the $\mathrm{\epsilon}$-regime where RMT is
applicable \cite{Bietenholz03}.

In our analysis, the would-be zero modes are first identified
and the configurations are classified according to their charge
indices obtained using the index theorem (\ref{eq_indextheorem}).
We should re-emphasize here that the indices
obtained by using different operators do not always agree
on a configuration by configuration basis but the charge
distributions are indistinguishable (see Fig.
\ref{fig_unimpglue_b5.85_v10_qdist} and
\ref{fig_impglue_b8.26_v10_qdist}). It should also be noted that
all configurations are used in our analysis. This is
different from some previous studies
\cite{Damgaard00,Follana04} where a small portion of the
ensemble was discarded. In these studies the topological
charge indices were also calculated using a direct discretization
of the continuum formula, $Q_{g}\sim \int \epsilon^{\mu\nu\rho\sigma}
F_{\mu\nu}F_{\rho\sigma}$. Only those configurations with
approximate integer value of $Q_{g}$ were included in the
analysis in Ref. \cite{Damgaard00}, and configurations for which
the charge index obtained by the index theorem is
different from $Q_{g}$ were excluded in Ref. \cite{Follana04}.

Results are shown in Fig. \ref{fig_rmt}. The solid
curves are predictions from
RMT. Because computing the spectra for the overlap operator
is much more expensive, the sizes of the ensembles are different:
1000 configurations for $(\mathrm{UFat7})\times\mathrm{Asq}$ and
HYP but only 400 for the overlap operator. This is the
reason that the overlap results appear to have 
poorer agreement with the predictions of RMT. Note that the agreement
is better with improved gauge fields. The results given here
contribute to the evidence that staggered fermions do feel gauge field topology,
provided that lattice artifacts are suppressed considerable,
as the agreement with the predictions of RMT is impressive.
They should be compared with previous studies \cite{gock99, Damgaard00} with
the unimproved staggered operator which appeared to indicate the presence of only 
a trivial topological sector.\\

%%%%%%%%%%%%%%%%%%%%%%%%%%%%%%%%%%%%%%%%
\section{\label{Conlcusion}Conclusion}
%%%%%%%%%%%%%%%%%%%%%%%%%%%%%%%%%%%%%%%%

In this project we studied numerically the spectral properties
of a variety of improved staggered Dirac operators. Four systematics have
been examined: i) improvement in the staggered operator, ii) improvement in
the gauge field action, iii) lattice spacing and iv) lattice volume.

It has been observed that the infrared eigenvalue spectrum
depends sensitively on the way in which the staggered
fermion operator is improved. On coarse lattices the unimproved
operator is insensitive to gauge field topology. As the level of
improvement increases, either on the operator itself or the
background gauge fields, eigenmodes with small eigenvalues and large
chirality appear.
These small eigenmodes can be identified as the chiral ``zero modes'' 
associated with the topology of the gauge fields. Sensitivity to the topology 
also increases as one approaches the continuum limit. This gives evidence
that lattice artifacts are responsible for the failure of the unimproved
staggered operator to reflect properly the gauge field topology on coarse 
lattices. Our results also show that a lattice spacing $a \lesssim 0.1fm$ is
enough for the Asqtad operator to have a correct response to the
topology with improved gauge fields. This spacing is of the order of
the lattice spacings used in present day state-of-the-act dynamical
simulations of QCD. On the other hand, the next level of improved staggered
operators, e.g., $\mathrm{UFat7}\times\mathrm{Asq}$, may be required
to produce configurations which describe the
correct continuum physics on coarser lattices.
We also observe that the topological
charge distribution is independent of which operator is used even
though the charge indices do not always agree on a configuration by 
configuration basis. However, the agreement increases with the level of 
improvement. A minimum physical volume of about $(1.2fm)^{4}$ seems to be
necessary in order for ``zero modes'' to show up and for sensitivity to 
topology to be established. This effect \cite{Bietenholz03,Zhang02}
was observed earlier for overlap fermions. As volume is increased the 
number of would-be ``zero modes'' increases as $\sqrt{V}$ but the number of 
low-lying non-chiral modes increases faster so some merging of tails of the
distributions takes place. Up to the $16^4$ volume considered here it is 
still possible to make a clear separation of would-be chiral modes from
non-chiral modes. The distribution of the
non-chiral modes is matched with the predictions of RMT.
The agreement is comparable to that obtained using overlap fermions.

Based on this, and other work \cite{Follana04,Wenger04}
done in the past year, one has strong evidence that, 
provided one uses improved staggered operators
and improved gauge fields, staggered fermions properly feel gauge
field topology.

\acknowledgments
We are very grateful to J. B. Zhang for providing the eigenvalue
solver for the overlap operator and H. D. Trottier for many useful
discussions. The computations were performed on 
facilities provided by WestGrid ({\tt{http://www.westgrid.ca/home.html}}).
This work was supported in part by the Natural
Sciences and Engineering Research Council of Canada.

%%%%%%%%%%%%%%%%%%%%%%%%%%%

\end{document}